\documentclass[aps,prl,twocolumn,groupedaddress]{revtex4}

\begin{document}

\renewcommand{\vec}[1]{{\mathbf #1}}

\title{Weak localization of short pulses in disordered waveguides}
\author{S.E. Skipetrov}
\email[]{Sergey.Skipetrov@grenoble.cnrs.fr}
\author{B.A. van Tiggelen}
\email[]{Bart.Van-Tiggelen@grenoble.cnrs.fr}
\affiliation{Laboratoire de Physique et Mod\'elisation des Milieux Condens\'es/CNRS,\\
Maison des Magist\`{e}res, Universit\'{e} Joseph Fourier, 38042
Grenoble, France}

\date{\today}

\begin{abstract}
We consider the phenomenon of weak localization of a short wave pulse in a
quasi-1D disordered waveguide. We show that the long-time decay
of the average transmission coefficient
is not purely exponential, in contradiction with predictions of the diffusion theory.
The diffusion theory breaks down completely for times exceeding the Heisenberg time.
We also study the survival probability of a quantum particle in a disordered waveguide and
compare our results with previous calculations using the super-symmetric nonlinear
sigma model.
\end{abstract}

\maketitle


Since the prediction of Anderson \cite{anderson58} that sufficiently strong disorder can block
propagation of waves and lead to localization of wave energy in space,
extensive efforts have been made to observe this `Anderson localization' for waves of various nature (Schr\"{o}dinger,
acoustic, electromagnetic, etc. waves) \cite{loc}. One particular experiment that makes localization evident is sending
a short wave pulse into a disordered medium and then observing its evolution in course of time.
If the disorder is weak, the wave will propagate out from the source by diffusion (at least at distances exceeding
the mean free path $\ell$) and its energy will be eventually distributed over the entire space, being negligible
at any given point.
If, in contrast, the disorder is sufficiently strong,
the scattering will prevent the wave from going away from the source and the energy of the wave will remain
localized within a volume of linear size $\xi$ around the source. The length scale $\xi$ is called the
localization length. The experiment to `detect' localization seems then trivial: just wait long enough and examine
which of the above scenarios is realized. In reality, however, the situation is complicated by the fact that
real disordered samples are finite in size and therefore, localized or not, the wave leaks out from the sample through
its boundaries, making the `final' (i.e. corresponding to very long times) state of the system identical for both weak and
strong disorder \footnote{Absorption of waves by the medium has the same effect.}. This implies that in order to
distinguish between diffuse and localized regimes, one has to analyze the leakage itself, and not only the final state of the system.
A correct analysis of the leakage requires careful treatment of boundary conditions. Indeed, the leakage happens
{\em through\/} the boundaries and hence the existence of the latter cannot be neglected.

The purpose of this paper is to present a relatively simple theoretical model that captures the main features
of the diffusion-localization transition correctly accounting for boundary conditions at the
surface of disordered sample. 
We then apply the model to describe the `weak localization' phenomenon, a precursor of
Anderson localization that can be observed in nominally diffusive disordered samples.
We limit ourselves to the case of the so-called `quasi-1D' experimental geometry
that is frequently encountered in experiments and is extensively studied theoretically. A quasi-1D sample is an open
cylindrical tube (waveguide) of length $L \gg \ell$, diameter $d$ such that $\lambda < d \leq \ell$
($\lambda$ is the wavelength),
and base surface $A = \pi d^2/4$.
The small diameter of the tube allows one to neglect the transverse variation of the average intensity of diffuse wave,
largely simplifying the derivations.
The tube has reflecting walls but open ends, and it is filled with a disordered medium (e.g. a mixture of transparent and scattering balls in the case of microwaves). A short wave pulse
is emitted at a position $z^{\prime}$ inside the tube at time $t = 0$ and the intensity of the wave is measured
at some other position $z$ at time $t$. In typical experiments, a wave is incident on the tube from outside and
the transmitted wave is measured. This corresponds to $z^{\prime} \simeq \ell$ (since $\ell$ is a typical distance needed
to convert the incident ballistic wave to the diffuse wave inside the sample) and $z = L$.
The following parameters are commonly used when light scattering in quasi-1D disordered waveguides is considered:
number of transverse modes $N = k^2 A/4 \pi \gg 1$ ($k = 2\pi/\lambda$) \footnote{We deal with scalar waves here. For vector (e.g. electromagnetic) waves, the number of transverse
modes is a factor of two larger due to two possible polarizations of the wave.}, dimensionless conductance
$g = (4/3) N \ell/L$ and localization length $\xi = (2/3) N \ell \gg \ell$.

We now consider a short wave pulse incident on the waveguide at $z = 0$ (in this case we set $z^{\prime} = \ell$)
or produced inside the waveguide (any $z^{\prime}$ between $0$ and $L$) at time $t = 0$.
To find the average transmission coefficient $T(t)$ of the tube measured at $z = L$, we apply the
self-consistent approach developed in Ref.\ \cite{skip04} and based on the
self-consistent theory of localization \cite{vw80}. It amounts to write a diffusion equation for
the intensity Green's function $C(z, z^{\prime}, \Omega)$:
\begin{eqnarray}
\left[ -i \Omega - \frac{d}{d z} D(z, \Omega) \frac{d}{d z} \right]
C(z, z^{\prime}, \Omega) = \delta(z - z^{\prime})
\label{selfcon1}
\end{eqnarray}
supplemented with a self-consistent equation for the position- and frequency-dependent diffusion coefficient:
\begin{eqnarray}
&&\frac{1}{D(z, \Omega)} = \frac{1}{D_0} + \frac{2}{\xi}
C(z, z, \Omega)
\label{selfcon2}
\end{eqnarray}
and with the boundary conditions:
\begin{eqnarray}
&&C(z, z^{\prime},\Omega) - z_0 \frac{D(z, \Omega)}{D_0} \frac{d}{d z}
C(z, z^{\prime},\Omega) = 0 \mbox{ at $z = 0$}
\label{bc1}
\\
&&C(z, z^{\prime},\Omega) + z_0 \frac{D(z, \Omega)}{D_0} \frac{d}{d z}
C(z, z^{\prime},\Omega) = 0 \mbox{ at $z = L\;\;\;\;\;$}
\label{bc2}
\end{eqnarray}
Here $D_0 = v \ell/3$ is the `bare' value of the diffusion coefficient ($v$ is the transport velocity)
and $z_0 \sim \ell$ is the so-called `extrapolation length' that allows one to account for reflection
of diffuse waves at the sample boundaries.
Equation (\ref{selfcon2}) describes the renormalization of the diffusion coefficient
due to the interference of time-reversed trajectories inside the disordered medium.
This renormalization has the same physical origin as the phenomenon of coherent backscattering
\cite{back}.
Note that the renormalized diffusion coefficient $D(z, \Omega)$ appears not only in the diffusion equation
(\ref{selfcon1}) but in the boundary conditions (\ref{bc1}, \ref{bc2}) as well.

If $g \rightarrow \infty$ (or, equivalently, if $L/\xi \rightarrow 0$),
waves propagate by diffusion and
the second term on the right-hand side of Eq.\ (\ref{selfcon2}) can be neglected.
We then have $D(z, \Omega) = D_0$, and by solving the diffusion equation (\ref{selfcon1}) with the boundary conditions
(\ref{bc1}, \ref{bc2}) we find 
\begin{eqnarray}
&&C_0(z, z^{\prime}, \Omega) = \frac{L}{\gamma D_0}
\nonumber \\
&&\times \left[ \sinh(\gamma z_{<}/L)
+ \gamma (z_0/L) \cosh(\gamma z_{<}/L) \right]
\nonumber \\
&&\times\left[ \sinh(\gamma (1 - z_{>}/L)) + \gamma (z_0/L) \cosh(\gamma (1 - z_{>}/L)) \right]
\nonumber \\
&&\times \left[ (1 + \gamma^2 (z_0/L)^2) \sinh \gamma + 2 \gamma (z_0/L) \cosh \gamma \right]^{-1}
\label{c0}
\end{eqnarray}
where we use the subscript `0' to denote the case of bare diffusion, $z_{<} = \min(z, z^{\prime})$,
$z_{>} = \max(z, z^{\prime})$, and $\gamma = (-i \Omega L^2/D_0)^{1/2}$.

We now consider very long waveguides ($L \gg \ell$) and neglect the extrapolation length
$z_0 \sim \ell \ll L$ in Eq.\ (\ref{c0}). This yields
\begin{eqnarray}
C_0(z, z^{\prime}, \Omega) &=& \frac{L}{D_0}
\frac{\sinh(\gamma z_{<}/L) \sinh\left[ \gamma (1 - z_{>}/L) \right]}{
\gamma \sinh \gamma}\;\;\;\;
\label{c0a}
\end{eqnarray}
The average transmission coefficient is then found as
\begin{eqnarray}
&&{\hat T}_0(\Omega) = -D_0 \frac{d}{d z} C_0(z = L, z^{\prime}, \Omega)
= \frac{\sinh(\gamma z^{\prime}/L)}{
\sinh \gamma}
\label{tomega}
\\
&&T_0(t) = \frac{1}{2 \pi} \int\limits_{-\infty}^{\infty}
{\hat T}_0(\Omega) \exp(-i \Omega t) d\Omega
\nonumber \\
&&= -i \sum\limits_{n = 1}^{\infty}
\mathrm{Res}\left[{\hat T}_0(\Omega) \exp(-i \Omega t), \Omega = -i n^2 \pi^2 D_0/L^2 \right]
\nonumber \\
&&= \frac{2 \pi D_0}{L^2}
\sum\limits_{n = 1}^{\infty} (-1)^{n+1} n \sin\left(n \pi z^{\prime}/L \right) \exp(-n^2 t/t_D)
\label{tt}
\end{eqnarray}
where $t_D = L^2/\pi^2 D_0$ is the time of wave diffusion through the waveguide.
To evaluate the integral in Eq.\ (\ref{tt}) we used the fact that the integrand
${\hat T}_0(\Omega) \exp(-i \Omega t)$ has simple poles
on the imaginary axis of the complex plane at $\Omega = -i n^2 \pi^2 D_0/L^2$ and applied the residue theorem.
Using Eqs.\ (\ref{tomega}) and (\ref{tt}) we can find the steady-state transmission coefficient:
\begin{eqnarray}
\int\limits_0^{\infty} T_0(t) dt \equiv {\hat T}_0(\Omega = 0) = \frac{z^{\prime}}{L}
\label{check}
\end{eqnarray}

At long times $t > t_D$ only the first term contributes appreciably into the sum of Eq.\ (\ref{tt})
and the average transmission coefficient decays exponentially with time:
$T_0(t) \simeq (2 \pi D_0/L^2) \sin(\pi z^{\prime}/L) \exp(-t/t_D)$.
This exponential decay is a hallmark of diffusion behavior and it is often used to test for
diffusion behavior of waves in experiments.
The diffusion constant $D_0$ can be extracted from the measured average transmission as
\begin{eqnarray}
D_0 = -\frac{L^2}{\pi^2} \frac{d}{d t} \ln T_0(t)
\label{d0}
\end{eqnarray}

However, recent experiments \cite{andrey03} show that even in the diffuse regime ($g \gg 1$) deviations from the simple exponential
decay of the average transmission coefficient can be detected.
This phenomenon is called weak localization and it has also an impact on the steady-state
($\Omega = 0$) transmission coefficient of a disordered waveguide. In the latter case, however,
the phenomenon is difficult to observe because one has to introduce some additional
factor (e.g. magnetic field) to brake the time-reversal symmetry, thus preventing the renormalization
of the diffusion constant, to be able to compare transmission coefficients in the presence and in the
absence of the time-reversal symmetry.
We now show how our theoretical model can be used to describe the weak localization
phenomenon for short wave pulses.

As far as $g$ remains much larger than unity, we can use $1/g$ as a small parameter and only keep
the terms of order $1/g$ (and drop the terms of order $1/g^2$, $1/g^3$, $\ldots$) in
Eqs.\ (\ref{selfcon1}) and (\ref{selfcon2}). This amounts to write
\begin{eqnarray}
D(z, \Omega) &=& D_0 + \frac{1}{g} D_1(z, \Omega)
\\
C(z, z^{\prime}, \Omega) &=& C_0(z, z^{\prime}, \Omega) + \frac{1}{g} C_1(z, z^{\prime}, \Omega)
\\
{\hat T}(\Omega) &=& {\hat T}_0(\Omega) + \frac{1}{g} {\hat T}_1(\Omega)
\\
T(t) &=& T_0(t) + \frac{1}{g} T_1(t)
\end{eqnarray}
Substituting these into Eqs.\ (\ref{selfcon1}) and (\ref{selfcon2}) and collecting the terms of order
$1/g$ we find
\begin{eqnarray}
&&D_1(z, \Omega) = -4 \frac{D_0^2}{L} C_0(z, z, \Omega)
\label{d1}
\\
&&\left[ -i \Omega - D_0 \frac{d^2}{d z^2} \right] C_1(z, z^{\prime}, \Omega) 
\nonumber \\
&&= \frac{d}{d z} \left[ D_1(z, \Omega) \frac{d}{d z} C_0(z, z^{\prime}, \Omega) \right]
\label{c1eq}
\end{eqnarray}
We now note that Eq.\ (\ref{c1eq}) with the delta-function $\delta(z-z^{\prime})$ on
the right-hand side is also obeyed by $C_0(z, z^{\prime}, \Omega)$.
$C_0(z, z^{\prime}, \Omega)$ is therefore the
Green's function of Eq.\ (\ref{c1eq}). The solution of the latter can then be written as
\begin{eqnarray}
&&C_1(z, z^{\prime}, \Omega) = -4 \frac{D_0^2}{L} \int\limits_0^L d z^{\prime\prime}
C_0(z, z^{\prime\prime}, \Omega) 
\nonumber \\
&&\times\frac{d}{d z^{\prime\prime}} \left[
C_0(z^{\prime\prime}, z^{\prime\prime}, \Omega) \frac{d}{d z^{\prime\prime}}
C_0(z^{\prime\prime}, z^{\prime}, \Omega) \right]
\label{c1int}
\end{eqnarray}
The integral in Eq.\ (\ref{c1int}) can be evaluated, and for $t > t_D$ we obtain
\begin{eqnarray}
&&T(t) = T_0(t)
\nonumber \\
&&\times \left\{
1 + \frac{1}{g} \left[\alpha_0(z^{\prime}) + \alpha_1(z^{\prime}) \frac{t}{t_D} + \alpha_2
\left( \frac{t}{t_D} \right)^2 \right] \right\}\;\;\;\;\;\;\;
\label{tt0}
\\
&&\frac{D(t)}{D_0} = 1 - \frac{1}{g} \left[ \alpha_1(z^{\prime}) + 2 \alpha_2 \frac{t}{t_D} \right]
\label{ddb}
\end{eqnarray}
where
\begin{eqnarray}
\alpha_0(z^{\prime}) &=& \frac{3}{4 \pi} \left( 1 - 3 \frac{z^{\prime}}{L} \right)
\sin\left( 2 \pi \frac{z^{\prime}}{L} \right)
\nonumber \\
&-& \frac{1}{4} \left( 1 - \frac{z^{\prime}}{L} \right) \left( 3 + \frac{z^{\prime}}{L} \right)
\label{a0}
\\
\alpha_1(z^{\prime}) &=& -\frac{3}{2 \pi^2} \cos\left( 2 \pi \frac{z^{\prime}}{L} \right)
\label{a1}
\\
\alpha_2 &=& \frac{1}{\pi^2}
\label{a2}
\end{eqnarray}
and $D(t)$ is defined by analogy with Eq.\ (\ref{d0}):
\begin{eqnarray}
D(t) = -\frac{L^2}{\pi^2} \frac{d}{d t} \ln T(t)
\label{dt}
\end{eqnarray}

Equations (\ref{tt0}) and (\ref{ddb}) are the main results of this paper. They lead us to the
following important conclusions. First, the decay of the average transmission coefficient $T(t)$ is
not purely exponential and hence, strictly
speaking, the propagation of waves in a disordered waveguide does not obey diffusion laws.
This is manifested by the time dependence of the diffusion coefficient (\ref{ddb}) which has
to be time-independent for the purely diffusive propagation. Second, the
diffusion model with a constant,
time-independent diffusion coefficient can be a good approximation only for short times
$t \ll g t_D$. It is easy to show that $t_H = g t_D$ is actually a characteristic time scale that is
known as the `Heisenberg time' in the field of quantum chaos. The Heisenberg time is the
typical time needed for a wave (or quantum particle) to visit the whole system (the whole
waveguide, in our case). After the Heisenberg time, the wave has to pass by those parts of the
sample that it has already visited before. The wave trajectory then crosses itself. Such
crossings are precisely the phenomenon that renormalizes the diffusion coefficient and that
is accounted for by the second term in Eq.\ (\ref{selfcon2}). Apparently, our theoretical model
does not allow us to treat the situations when such crossings become dominant.
We cannot therefore treat times exceeding the Heisenberg time. Another way to understand this
limitation of our theoretical approach is to note that the Heisenberg time is also the
inverse of the typical spacing between the modes of a closed sample. For times exceeding 
$t_H$ the mode structure of the wave field becomes important (i.e. one resolves individual
modes). This mode structure is not included in our model and hence we cannot treat this
regime correctly.

It is worth-noting that $\alpha_0(z^{\prime})$ and $\alpha_1(z^{\prime})$
in Eqs.\ (\ref{tt0}) and (\ref{ddb}) depend on the position $z^{\prime}$ of the source
of waves and can even change sign when
the source position is changed. For example, when the wave is incident on the waveguide from outside
(as in the experiments of Chabanov {\em et al.} \cite{andrey03}), we set $z^{\prime}\simeq \ell$
and find $\alpha_0 = -3/4$ and $\alpha_1 = -3/2 \pi^2$ for $L \gg \ell$. On the other hand,
if the source is placed in the middle of the waveguide, we find
$\alpha_0 = -7/16$ and $\alpha_1 = 3/2 \pi^2$. In contrast, the value of
$\alpha_2$ is not sensitive to the source position. This illustrates that for long times
the waves explore the whole sample and have only a limited memory about the position of
their source. 

The decrease of the diffusion coefficient with time predicted by Eq.\ (\ref{ddb})
has been recently observed in microwave
experiments by Chabanov {\em et al.} \cite{andrey03}. Out theoretical model allows a reasonably
good description of the experimental results \cite{skip04}. It is also worthwhile to note that
a result similar to Eq.\ (\ref{tt0}) has been obtained by Mirlin \cite{mirlin00} who studied the
`survival probability' $P(t)$ of a quantum particle in a quasi-1D disordered waveguide using the
super-symmetric nonlinear sigma model and found
\begin{eqnarray}
-\ln P(t) = \frac{t}{t_D} \left( 1 - \frac{1}{\pi^2 g} \frac{t}{t_D} \right)
\label{survival}
\end{eqnarray}
for orthogonal symmetry ($\beta = 1$), corresponding to the case of preserved
time-reversal symmetry considered here.
To calculate $P(t)$ in the framework of our self-consistent approach we note that the
reason for the survival probability to differ from $1$ is the leakage of wave energy
through sample boundaries.
This leakage is quantified by the transmission coefficient $T(t)$ and we therefore find
\begin{eqnarray}
\frac{d P(t)}{dt} = - 2 {\bar T(t)}
\label{mysurvival}
\end{eqnarray}
where the bar denotes averaging over the position of the source $z^{\prime}$ and the numerical
factor 2 is the number of (identical) open boundaries.
For bare diffusion ($g \rightarrow \infty$) we substitute Eq.\ (\ref{tt}) into
Eq.\ (\ref{mysurvival}) and using the `initial' condition $P(0) = 1$ find
\begin{eqnarray}
&&P_0(t) = \frac{8}{\pi^2} \exp\left(-\frac{t}{t_D} \right) \mbox{ and }
\nonumber \\
&&-\ln P_0(t) =  \frac{t}{t_D} -\ln\left( \frac{8}{\pi^2} \right)
\label{p0}
\end{eqnarray}
for $t > t_D$. As expected, the decay of $P_0(t)$ is purely exponential.
Note that instead of using the condition $P(0) = 1$ we could arrive at the same
result (\ref{p0}) by requiring $P(\infty) = 0$.

Consider now finite but large $g$. Substituting Eq.\ (\ref{tt0}) into
Eq.\ (\ref{mysurvival}) and performing the integration yields $P(t)$ up to an additive constant.
The condition $P(0) = 1$ cannot
be used to determine this constant (as for $g \rightarrow \infty$) because Eq.\ (\ref{tt0})
only holds for $t > t_D$ and does not apply for $t = 0$. We therefore use the condition
$P(\infty) = 0$ to determine the additive constant and obtain
\begin{eqnarray}
P(t) &=& \frac{8}{\pi^2} \exp\left(-\frac{t}{t_D} \right)
\nonumber \\
&\times&
\left\{ 1 + \frac{1}{g} \left[ {\tilde \alpha_0} + {\tilde \alpha_1}
\frac{t}{t_D} + {\tilde \alpha_2} \left( \frac{t}{t_D} \right)^2 \right] \right\}
\label{pp}
\end{eqnarray}
where ${\tilde \alpha_0} = 3/\pi^2 - 3/8$, ${\tilde \alpha_1} = 5/2 \pi^2$, and
${\tilde \alpha_2} = 1/\pi^2$. This is equivalent to
\begin{eqnarray}
-\ln P(t) &=& \frac{t}{t_D}
\left( 1 - \frac{5}{2 \pi^2 g} - \frac{1}{\pi^2 g} \frac{t}{t_D} \right)
\nonumber \\
&-&\left[ \ln\left( \frac{8}{\pi^2} \right) + \frac{1}{g} \left( \frac{3}{\pi^2} -\frac{3}{8} \right)
\right] 
\label{pp2}
\end{eqnarray} 

If we drop the time-independent terms in the square brackets [which seems to be done
in Eq.\ (\ref{survival})], the resemblance between Eqs.\ (\ref{survival}) and
(\ref{pp2}) becomes evident. Indeed, the numerical coefficient $-1/\pi^2 g$ in front of
$(t/t_D)^2$ is {\em the same\/} in both cases. On the other hand, the coefficients in front
of $t/t_D$ differ by $5/2\pi^2 g$. It is interesting to note that the coefficients
in front of $(t/t_D)^2$ and  $t/t_D$ do not depend on the exact condition
[$P(\infty)=0$, $P(0) = 1$ or
$P(t \rightarrow 0) = P_0(t \rightarrow 0$)] that we employ to determine the additive constant
when integrating Eq.\ (\ref{mysurvival}).

In conclusion, we have presented a theoretical model that enables us to describe the phenomenon
of localization for short wave pulses in disordered media, correctly accounting for boundary
conditions on the sample surface. The model is based on a
diffusion equation with a self-consistently renormalized diffusion coefficient.
Using this model, we calculate the time-dependent transmission of waves through
a disordered quasi-1D waveguide and the survival probability of a wave (or quantum particle) in
the waveguide in the diffuse regime (dimensionless conductance much larger than unity).
We show that the decay of these quantities with time is not purely exponential,
even for nominally diffusive samples, and slows down due to the phenomenon of weak localization.
We compare our result for the survival probability with that obtained previously using
a different approach, and show that the two results are similar, but not completely identical.


\end{document}